\relax
\documentstyle[epsf,prl,twocolumn,aps]{revtex}

\newcommand{\be}{\begin{equation}}
\newcommand{\ee}{\end{equation}}
\newcommand{\bea}{\begin{eqnarray}}
\newcommand{\eea}{\end{eqnarray}}

\newcommand{\p}{\partial}
\newcommand{\ri}{{\rm i}}
\newcommand{\re}{{\rm e}}
\newcommand{\rd}{{\rm d}}
\newcommand \varep {\varepsilon}
\begin {document}
\bibliographystyle {plain}
%\tableofcontents 
%\setlength {\textwidth}{14.5cm
%\setlength {\textwidth}{14.5cm
%\setlength {\textwidth}{14.5cm\pagestyle{nonempty}

%***********    This is for two columns *******************************
\twocolumn[\hsize\textwidth\columnwidth\hsize\csname @twocolumnfalse\endcsname
%********************************

\title{ Induced Luttinger Liquid Behaviour in an Exactly Solvable Model of
Stripes}
\author{Catherine P\'epin and  Alexei M. Tsvelik}
\address{Department of Physics, University of Oxford, 1 Keble Road, Oxford, 
OX1 3NP, UK}
\date{\today }
\maketitle

\begin{abstract}
\par
We study an exactly solvable model describing 
 a stripe   consisting of a Toda array of $N$ anharmonic elastic chains
 sandwiched between  two
conducting chains.  It is shown that 
the presence of a 
charge  on  one chain generates a gapless excitation branch (Luttinger
 liquid) on the
other and  leads to increase in the phonon frequencies.

\end{abstract}
%cond-mat/9612014
\pacs{75.10 Jm, 75.40 Gb. } 
\sloppy

%\newpage
% ***********    This is for two columns *******************************
\vskip2pc]
%********************************

\par

  During recent several years there have been rather intense interest
 in systems which combine one-dimensional and two- or
 three-dimensional features. As examples of such systems one can
 mention domain wall (or  stripe)  formations in doped Mott insulators
 \cite{emery} and quasi-one-dimensional 
systems with electron-phonon interactions. The capital fact about
 phonon  
systems is that phonons
 are {\it never} one-dimensional. Thus in  both cases electrons
 residing on 
 one-dimensional chains are immersed in a two- or three-dimensional
 {\it active} environment. 
 This feature makes such problems  very challenging from the
 theoretical  point of view 
because on one
  hand the low dimensionality leads to the enhancement of interactions
 thus requiring non-perturbative approach,  and
 on the other hand the multi-chain structure fits rather awkwardly
 with available non-perturbative techniques. Usually  theorists
 resort to a hybrid treatment incorporating  features of exact
 solution for individual chains and RPA (Random Phase Approximation)
 for interchain interactions \cite{essler}. 

  From the qualitative point of view there are several  interesting
 problems concerning the systems in question. One is how conducting
 regions 
 affect each other and another is how changes in the conducting region 
 affect the
 active medium in which it exists.

\begin{figure}
% ********   This is for two columns
\epsfysize=1.5in 
% ***********For one column  ********************
%\epsfxsize=1.5in 
% ***********************************8
\centerline{\epsfbox{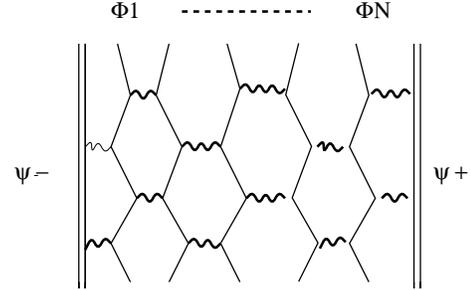}}
\vskip 0.1truein
\protect\caption{The  system under consideration: two conducting
regions  coupled by the  Toda array.}
\label{figtoda}
\end{figure}

Here we discuss these problems using as an example an exactly solvable
model
of a static stripe  
suggested by Fateev~\cite{fateev1}. A remarkable property of this model is that 
 it exists in  two
 different representations describing situations with 
different physics. One representation is fermionic and the other one
 is bosonic and they are related by a duality transformation. 
In both its  incarnations  the model  describes
 two species of 
(1+1)-dimensional charged particles located on different sides of an insulating
 stripe and interacting via its elastic modes (see Fig.~\ref{figtoda}). 
The latter ones are
 anharmonic optical phonons described by the so-called Toda array. 
The Lagrangian  density 
has the following form:
\be
{\cal L}(\beta, N) = {\cal L}_{Toda} + {\cal L}_{edge}
\ee
\bea
{\cal L}_{Toda} &=& \frac{1}{2}\sum_{n = 1}^N(\p_\mu\phi_n)^2  
- (m/\beta)^2\sum_{n = 1}^{N -
1}\re^{\beta(\phi_{n+1} - \phi_{n})}\label{Toda}
\eea
In its fermionic incarnation the edge Lagrangian  is given by
\bea
{\cal L}_{edge}^{(f)} &=&  \sum_{s =
1,2}\left[\ri\bar\psi_s\gamma_{\mu}\p_{\mu}\psi_s + 
\frac{\pi g}{2} (\bar\psi_s\gamma_{\mu}\psi_s)^2\right] 
- 
m\bar\psi_1\psi_1\re^{-\beta\phi_1} \nonumber\\
&-&
m\bar\psi_2\psi_2\re^{\beta\phi_N} -(m/\beta)^2\left(\re^{-
2\beta\phi_1} + \re^{2\beta\phi_N}\right)\label{fermionic}
\eea
where $g  = \beta^2/(4\pi + \beta^2)$. It can be interpreted as a
model of two charge density waves coupled through the Toda array.

 The bosonic representation is more intriguing because of its 
 connotations with the hypothetic stripe formations in doped Mott
 insulators. The edge Lagrangian density is 
\begin{eqnarray}
{\cal L}_{edge}^{(b)} &=&\frac{1}{2}\sum_{s = 1}^2\frac{\partial
_{\mu}\Delta_s ^{*}\partial _{\mu}\Delta_s }{1+(\beta /2)^{2}\Delta_s
^{*}\Delta_s }\nonumber\\
&-&
\frac{m^{2}}{2}(|\Delta_1 |^{2} + 2){\rm e}^{- \beta \phi _{1}} -
\frac{m^{2}}{2}(|\Delta_2 |^{2} + 2){\rm e}^{\beta \phi _{N}}
 \label{bosonic}
\end{eqnarray}
where $\Delta_s$ are complex bosonic fields which can be interpreted
as superconducting order parameters residing on the edges of the
stripe. 

 The duality transformation relating fermionic to the bosonic
representation is 
\be
L^{(f)}(\beta,N) = L^{(b)}(4\pi/\beta, N - 2)
\ee
which also corresponds to $g \rightarrow 1 -g$. Therefore the strong
coupling limit of one model is related to the weak coupling limit of
another one.

The model has U(1)$\times$U(1) symmetry and the corresponding charges
are given by 
\bea
Q_s = \int \rd x\bar\psi_s\gamma_0\psi_s = -\frac{\ri}{2}\int \rd
x\frac{(\Delta^*_s\p_0\Delta_s - \Delta_s\p_0\Delta^*_s)}{[1 + (\beta/2)^2|\Delta_s|^2]}
\eea 
One can modify the Hamiltonian introducing chemical
potentials coupled to the charges $H \rightarrow 
H - h_+Q_+ - h_-Q_-$ which would
correspond to populating the edges of the stripe. In this publication
we shall be particularly interested in the situation when one chemical
potential is much greater than another. 

 In  \cite{fateev1} Fateev gave a proof 
 of  integrability of the above models and suggested   exact two-body
 $S$-matrices.  At  zero  chemical potentials  
all excitations have gaps. The particles living on the edges may be
considered as  fundamental ones,  excitations of the Toda array being
their bound states. There are $N - 1$   bound states with the masses 
\be
M_j = 2M\sin(\pi j/2\lambda),  (j = 1,2, ... N - 1),  \lambda = N - g 
\label{mass}
\ee
where $M = M(m, \beta)$ is the mass of the fundamental particle. All
bound states are neutral. 
The case $N = 1$ where there are no bound states  has been
thoroughly studied in \cite{fateev2}.

 Each fundamental particle is labeled by two indices corresponding to
the two U(1) groups (it is either soliton or antisoliton and it
belongs to one chain or another). 
The two-body scattering matrix of the fundamental 
particles is given by a tensor product of two sine-Gordon
$S$-matrices multiplied by a CDD-factor responsible for calcellation
of double poles. Since the Bethe ansatz equations for the sine-Gordon
model is well known, it is quite straightforward to write down such
equations for the model in question:

\bea
\re^{-\ri M\sinh\theta_j L} & =&  \prod_{k \neq j}S_0(\theta_j -
\theta_k)  \times  \nonumber \\
& & \prod_{a = 1}^{m_1}
e_1(\theta_j - u_a) \prod_{b = 1}^{m_2}e_1(\theta_j - v_b)\label{ba1} \\
\prod_{j = 1}^ne_1(\theta_j - u_a) & = &
\prod_{b = 1}^{m_1}e_2(u_b - u_a)\label{ba2} \\
\prod_{j = 1}^ne_1(\theta_j - v_a) & = & 
\prod_{b = 1}^{m_2}e_2(v_b - v_a)\label{ba3}
\eea
where $
e_n(x) =  \sinh\lambda(x +
\ri n\pi/2) / \sinh\lambda(x - \ri n\pi/2)
$, the numbers 
$m_{1,2} = n/2 - Q_{1,2}$ and 
\[
S_0(\theta) = \re^{2\ri\delta_{\lambda}(\theta)}\prod_{a = 1}^{N - 1}\frac{\sinh\theta -
\ri\sin(\pi a/\lambda)}{\sinh\theta + 
\ri\sin(\pi a/\lambda)} ~~~~~~~ {\mbox with}
\]
  
\[
\delta_{\lambda}(\theta) = \int_0^{\infty}
\rd\omega\frac{\sin(\omega\theta)}{\omega}\frac{\sinh[\pi\omega(\lambda^{-1}
- 1)/2]}{\cosh(\pi\omega/2)\sinh(\pi\omega/2\lambda)}
\]

\begin{figure}[tb]
\twocolumn[\hsize\textwidth\columnwidth\hsize\csname
@twocolumnfalse\endcsname
\epsfxsize=5.5truein 
\epsfysize= 2.3truein
\centerline{\epsfbox{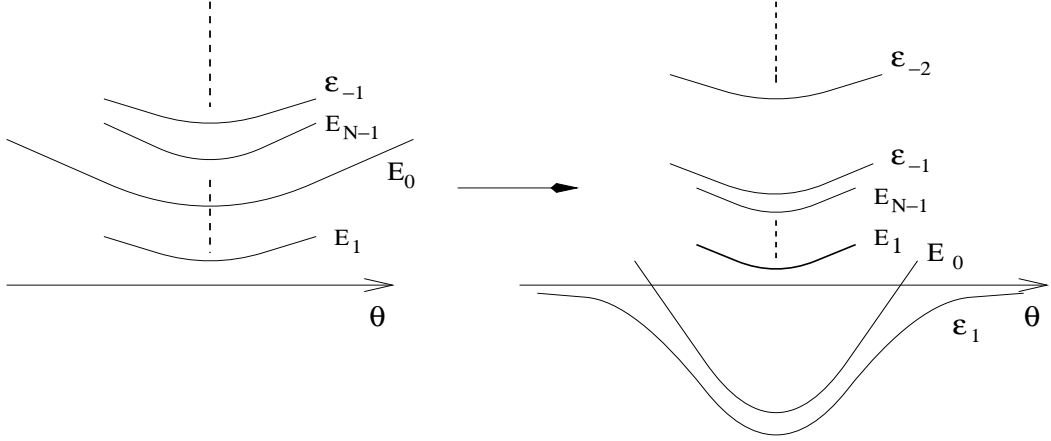}}
\vskip 0.1truein
\protect\caption{Energies  $\epsilon_n(\theta)$ before (left) and after (right)
application of a chemical potential $h_-$ on the left wire. Only those
 modes  are shown whose energies  do not vanish at $T = 0$; the chemical
 potential creates large gaps $\sim (n - 1)h_-$ for $\epsilon_{-n}$ (n > 2)
 and smaller gaps $\sim \Delta_{ph}$ (\ref{newgap}) for the phonon
 modes  
$E_j$ (j = 1, ... N -1) and $\epsilon_{-1}$. The mode  $\epsilon_1$
 becomes gapless. 
 }
\label{figlevels}
\vskip 0.2truein]
\end{figure}

  We claim that 
that creating a charge on  one shore   of the stripe
 one generates  a band of soft collective modes on the other
shore. This statement  can
be understood without detailed calculations which we leave for the
extended publication, just using a qualitative analysis of the Bethe ansatz
equations (\ref{ba1}, \ref{ba2},\ref{ba3}) and will be illustrated on figure~\ref{figlevels}. As a matter of fact,
similar  
effect may  occur  in
all integrable systems where the $S$-matrix is a tensor product (here 
Principal Chiral Fields may serve as a good example - see, for
example,  \cite{wiegmann}). In all these
cases the chemical potential coupled to one of the conserved charges
generates  not only its own charge, but also induces a finite
susceptibility with respect to the other chemical potential. Thus
 the second subsystem becomes gapless. We call this effect the
induced Luttinger liquid.

 Notice that  Eqs. (\ref{ba2},\ref{ba3}) 
for $u_a$, $v_a$ look very
similar to 
BA equations for the spin-1/2 XXZ spin chain (see, for example,
 \cite{takahashi})
with the anisotropy related to the parameter $g$.  
In the presence of a 
large chemical potential $h_- \ll M$ applied on the left shore, the density of particles in the
Fermi sea  
 $n/L$ is finite and the real rapidities $\theta$ in  
Eqs.(\ref{ba1},\ref{ba2},\ref{ba3}) with energie $E_0$ are densely
distributed over 
 some finite interval  $[-B,B]$. The 
excitation energies $ \varep_{-n}$ of the
$u$-particles are pushed up by the chemical potential.
The excitation energies $\epsilon_{n}$ of the $v$- particles have their minima at
 $|v| \rightarrow \infty$ and therefore at $T \ll h_-$
 (the Fermi energy of the $\theta$-particles) they are located very
 far away in the momentum space from the Fermi momenta of the 
$\theta$-particles. 
As all energies $\epsilon_{-n}$ are gapped,the rapidities $u$ drop out of consideration and in the
equations for $v$ one can neglect a feedback between $v$'s  and
 $\theta$'s 
and replace $\theta$'s by their ground state
values. 
Furthermore, at large  values of  
$|v| \ll B$ which are important at low energies in these equations 
one can neglect 
$\theta$'s in comparison with $v$'s. Then 
Eqs.(\ref{ba3}) become  exactly like the Bethe ansatz equations for 
 the spin-1/2 XXZ spin chain  with the 
total number of sites $n$ equal to the number of
$\theta$'s in the ground state:
\bea
H = J\sum_j^n[S_j^xS_{j + 1}^x + S_j^yS_{j + 1}^y + \cos(\pi g) S_j^zS_{j +
1}^z] \label{xxz}
\eea
where we set $0 < g < 1$ using the fact that 
 the region $g < 0$ is mapped onto
$g > 0$ by the transformation $1 - g \rightarrow -g$. 
 Since all these arguments 
work only for small energies
where the spin chain is equivalent to the Luttinger liquid, one can say
that the $v$-subsystem describes a  spinless Luttinger liquid 
with the Luttinger
parameter 
\be
K = [2(1 - g)]^{-1}
\ee
 We emphasise that this value of $K$
is independent of number of chains or the number of particles in the
Fermi sea.

The corresponding velocity $V_s$ 
however, does depend on $N$ and the
 number of particles.
One can show that
$V_s$ vanishes at $g
 \rightarrow 0,1$ indicating that the collective band described here
 is generated by the interactions.

 Now we shall briefly outline the technical details leading to the
 above conclusions.  Following the standard procedure we can  
 derive the Thermodynamic Bethe Ansatz (TBA) equations
 corresponding to Eqs.(\ref{ba1},\ref{ba2},\ref{ba3}). To make the
 problem simplier we use the fact that the above Bethe ansatz
 equations are invariant under the transformation 
\bea
- g &=& 1 - g^\prime, N = N^\prime - 1
 \nonumber\\
u_a\lambda &=&  \tilde u_a\lambda +
\ri\pi/2, ~~v_a\lambda =  \tilde v_a\lambda +
\ri\pi/2
\eea
Due to the periodicity of the phase factors in
 Eqs.(\ref{ba1},\ref{ba2},\ref{ba3}) one can make a replacement 
\[
\frac{\sinh\lambda(u + \ri n\pi/2)}{\sinh\lambda(u - \ri n\pi/2)}
 \rightarrow \frac{\sinh(\lambda u - \ri\pi n g/2)}{\sinh(\lambda u + \ri
\pi ng/2)}
\]
These two facts allow us to restrict our consideration by  the   region
$1 > |g| > 1/2$. To further simplify our presentation  we shall give
explicit expressions only for the case 
 $g = 1/\nu > 0$  with  $\nu \geq 2$
 being  an integer number where  classification of solutions  is simplier. 
The free energy is  then given by  
\bea
F/L = - \frac{1}{2\pi}\sum_{j = 0}^{N- 1} M_j\int d\theta \cosh\theta
\ln[1 + \re^{-E_j(\theta)/T}]
\eea
where  $M_0 = M$ and $M_j$ are given by
Eq.(\ref{mass}).

 The TBA equations for the bound states are   
\bea
& T& \ln[1 + \re^{E_j(\theta)/T}] - 
G_{jk}*T\ln[1 + \re^{-E_k(\theta)/T}] = M_j\cosh \theta \nonumber\\
& + &
G_{j0}*T\ln[1 + \re^{-E_0(\theta)/T}], ~~(j,k = 1, ... N - 1) \label{tbab}
\eea
where  the  star denotes convolution 
\[
f*g(\theta) = \int_{-\infty}^{\infty} d\theta'f(\theta -
\theta')g(\theta')
\]
In this publication we do not need explicit expressions for the
kernels $G_{jk}, G_{0j}$; the important 
features of Eqs.(\ref{tbab}) being that $G_{j0}
> 0$ and the energies $E_j$ are directly coupled only to $E_0$. The latter property
is related to the fact that bound states are neutral with respect to
both U(1) groups.

\begin{figure}
% ********   This is for two columns
\epsfysize=1.0in 
\epsfxsize=3.0in
% ***********For one column  ********************
%\epsfxsize=7.0in 
% ***********************************8
\centerline{\epsfbox{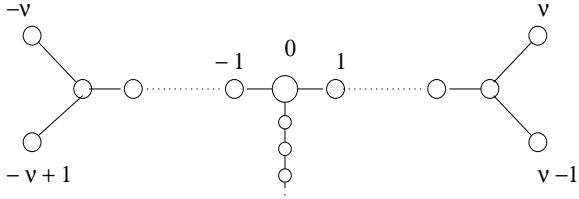}}
\vskip 0.1truein
\protect\caption{The incident diagram for the TBA equations given in
the text. The circles correspond to energies and links to
nonvanishing kernels. The circles on the vertical line correspond to
the bound states. } 

\label{fig1}
\end{figure}

When the chemical potentials $h_-$ and $h_+$ are applied respectively
to the left and right edges, the full  system of TBA is given by

\bea
& T& \ln[1 + \re^{E_0(\theta)/T}] - K*T\ln[1 + \re^{-E_0(\theta)/T}]
\nonumber \\
&=& M\cosh\theta - (h_+ + h_-)/2 
+ G_{0j}*T\ln[1 + \re^{-E_j(\theta)/T}]\nonumber\\
 & - &  T \sum_{n = 1}^{\nu - 1}a_{n}*\ln[1 + \re^{-\epsilon_{-n}(\theta)/T}]
- Ta_{\nu -1}*\ln[1 + \re^{-\epsilon_{-\nu}(\theta)/T}] \nonumber \\
 & - & Ts*\ln[1 + \re^{\epsilon_1(\theta)/T}] \label{tba1}
\eea

equations for $\epsilon_n$ and $\epsilon_{-n}$ are

\bea
\epsilon_{\pm\nu}  & =& h_{\pm}\nu/2 - 
s*T\ln[1 + \re^{\epsilon_{\pm( \nu -2)/T}}] \label{tba2} \\
\epsilon_{\pm n}&  = & \delta_{|n|,\nu -1}h_{\pm}\nu/2 + s*T\ln[1 +
\re^{-\epsilon_{ \pm n
- 1}/T}][1 + \re^{\epsilon_{\pm n + 1}/T}]  \nonumber\\
& + & \delta_{\nu - 2,|n|}s* T\ln[1 + \re^{- \epsilon_{\pm \nu}/T}]
\nonumber \\
& - &
\delta_{|n|,1}s*T\ln[1 + \re^{-E_{0}/T}] \label{tba3}
\eea
where 
\[
a_n(\omega)   = \frac{\sinh[\pi(1 -
 n/\nu)\omega/2\lambda]}{\sinh[\pi\omega/2\lambda]}, s(\omega) = [2\cosh(\pi\omega/2\lambda\nu)]^{-1}
\]
\[
K(\omega) =   \frac{\sinh[\pi\omega(1 - g)/2\lambda]\cosh[\pi\omega(\lambda +
2g)/2\lambda]}{2\cosh\pi\omega/2\cosh(g\pi\omega/2\lambda)\sinh(\pi\omega/2\lambda)}
\]
The general structure of the TBA can be illustrated by the incident
diagram drawn on Fig.~\ref{fig1}.  From these equations it is manifest  that at $h_+ = 0$ and $h_-
 > M$ when $E_0$ has a negative part, all $\epsilon_{-n}$ are positive except, possibly
    $\epsilon_{-1}$  ( see Fig.~\ref{figlevels}).

 From Eq.(\ref{tba1}) we see that at  $h_- > M$ the energy 
$E_0 $ is negative   on some interval $[-
B, B]$. From Eq.(\ref{tbab}) it is clear  that $E_j$'s are pushed
further up. As follows from (\ref{tba1}) at $T = 0$ 
\be
E_0 \approx K^{-1}*(M\cosh\theta - h_-/2)
\ee
Substituting this into Eq.(\ref{tbab})  we find that the mass gaps for
 bound states are of order of 
\be
\Delta_{ph} \sim M(h_-/M)^\eta, ~~\eta = g/(N + g) \label{newgap}
\ee 
 One can also check that all $\epsilon_n$ except of 
$\epsilon_{-1}$ still have a gap of order of $h_-$ and $\epsilon_{-1}$
 has a much smaller gap $\sim \Delta_{ph}$. 
  Thus at $h_- \gg M$ there are two temperature scales in the problem:
$h_-$ itself which play a role of the Fermi energy; at temperatures $T \ll h_-$ one can omit $\epsilon_{-n}$'s
 from Eq.(\ref{tba1}). The other scale is the much smaller one - it is
the phonon frequency scale 
$\Delta_{ph}$. 
It also follows from Eqs.(\ref{tba3}) that  all $\epsilon_n$'s with $n >1$ are
 positive and of order of $T$ and $\epsilon_1$ is purely negative in
 the limit $T = 0$. Therefore the term with $\epsilon_1$ in the right
 hand side of Eq.(\ref{tba1}) gives no contribution at $T = 0$.

>From now on we shall  ignore bound states assuming that
  $T \ll \Delta_{ph}$. 
At  such temperatures we can neglect temperature dependence of
$B$. We are interested in the region where 
  all $\epsilon_n$ are
small.  This corresponds to  $|\theta| \rightarrow \infty$ which are 
 far from the Fermi surface of 
$E_0(\theta)$. Therefore   we can 
replace $- T\ln[1 + \re^{-E_{0}(\theta)/T}]$ in 
the right-hand side of Eqs.(\ref{tba2},\ref{tba3}) by its value at zero
temperature 
$\epsilon^{(-)}(\theta) = E_0(\theta)\Theta[-
E_0(\theta)]$. For $\theta \gg B$ the driving term can be replaced by
its asymptotics:
\[
s*\epsilon^{(-)}(\theta) \approx s(\theta)\int_{-B}^B
dy E_{0}(y)\cosh(\lambda y/g)
\]
After this substitution, the TBA equations (\ref{tba2},\ref{tba3}) 
coincide with the low temperature limit of the TBA
equations for the anisotropic spin-1/2 Heisenberg magnet (see
Eq.(\ref{xxz}) above). 
 
 In order to obtain the free energy  at low $T$ we 
 isolate contributions  which vanish at $T = 0$. This is achieved
 using  the standard TBA machinery;  the free energy is represented as
 a sum of two parts corresponding to contributions from two gapless modes:
\[
F/L = f_1 + f_2 ~~~~~~~~~~ {\mbox with}  
\]
\bea
f_1 & =& - T\int_{-\infty}^{\infty}[\sigma^{(+)}(\theta) +
\sigma^{(-)}(\theta)]\ln(1 + \re^{- |E_0(\theta)|/T})d\theta \nonumber\\
& \approx & - \pi^2 T^2/3V_c ~~~~~~{\mbox and} \\
f_2 & = & - T\int d\theta s*\sigma^{-}(\theta)\ln[1 +
\re^{\epsilon_1(\theta)/T}]  \nonumber \\
& \approx & - \pi^2 T^2/3V_s  \ .
\eea
This describes independant gapless modes on the left edge ($f_1$) as
well as the induced mode on the right shore ($f_2$).
The function  $\sigma(\theta)$ satisfies 
\[
\sigma^{(+)}(\theta) + \int_{-B}^B d\theta^\prime  K(\theta -
\theta^\prime)\sigma^{(-)}(\theta^\prime) = \frac{M}{2\pi}\cosh\theta \label{sigma}
\]
where $\sigma^{(+)}(\theta) = 0$ at $|\theta| < B$ and
$\sigma^{(-)}(\theta) = 0$ at $|\theta| > B$. The latter function is
the ground state density of real $\theta$'s. 

For very large chemical potential $h_-$ ( $B \rightarrow \infty$), we
 find the velocities $V_c = 1$ and the velocity of the inducted mode 
$V_s = g/(N - g)$. Therefore
 the inducted  gapless band collapses at $g \rightarrow 0$ as it
 should. The similar thing  happens at $g \rightarrow 1$. 

  In conclusion,
  charging of one edge of the stripe creates a band of gapless
  excitations on the other edge and stiffens the phonon modes
  inside. Their energies, however, remain much smaller than the Fermi
  energy which means that the system in question has an intermediate
  energy scales where one should expect nontrivial
  crossovers.

 We are grateful to P. Coleman and F. Essler for
 valuable criticisms  
 and  to V. A. Fateev for 
hospitality  and 
interest to the work.

%\newpage

\end{document}